\documentclass{sf2a-conf2023}
\usepackage{graphicx}
\usepackage{hyperref}
\usepackage[]{natbib}  
\usepackage{epstopdf}

\def\BibTeX{{\rm B\kern-.05em{\sc i\kern-.025em b}\kern-.08em
    T\kern-.1667em\lower.7ex\hbox{E}\kern-.125emX}}
\bibpunct{(}{)}{;}{a}{}{,}  

\begin{document}

\TitreGlobal{SF2A 2023}


\title{Measuring stellar rotation and activity with PLATO}

\author{S. N. Breton}\address{INAF – Osservatorio Astrofisico di Catania, Via S. Sofia, 78, 95123 Catania, Italy}

\author{A. F. Lanza$^1$}

\author{S. Messina$^1$}

\author{R. A. García}\address{Universit\'e Paris-Saclay, Universit\'e Paris Cit\'e, CEA, CNRS, AIM, F-91191, Gif-sur-Yvette, France}

\author{S. Mathur}\address{Instituto de Astrof\'{i}sica de Canarias, 38205 La Laguna, Tenerife, Spain \\ Departamento de Astrof\'{i}sica, Universidad de La Laguna (ULL), 38206 La Laguna, Tenerife, Spain}

\author{A. R. G. Santos}\address{Instituto de Astrof\'isica e Ci\^encias do Espa\c{c}o, Universidade do Porto, CAUP, Rua das Estrelas, PT4150-762 Porto, Portugal}

\author{L. Bugnet}\address{Institute of Science and Technology Austria (IST Austria), Am Campus 1, Klosterneuburg, Austria}

\author{E. Corsaro$^1$}

\author{I. Pagano$^1$}




\setcounter{page}{237}


\maketitle


\begin{abstract}
Due to be launched late 2026, the PLATO mission will bring the study of main-sequence solar-type and low-mass stars into a new era. In particular, PLATO will provide the community with a stellar sample with solar-type oscillations and activity-induced brightness modulation of unequalled size. We present here the main features of the analysis module that will be dedicated to measure stellar surface rotation and activity in the PLATO Stellar Analysis System. 
\end{abstract}



\section{Introduction}

The Planetary Transit and Oscillations of stars \citep[PLATO,][]{Rauer2014} mission will provide high-cadence photometric light curves for hundreds of thousand of solar-type stars. The Module for Stellar Astrophysics number 4 (MSAP4), currently under development, is the PLATO pipeline component dedicated to analyse stellar surface rotation and surface magnetic activity including magnetic cycles in the light curves acquired by the mission.  

\section{The MSAP4 module}

\subsection{General framework}

\begin{figure}[ht!]
    \centering
    \includegraphics[width=0.9\textwidth]{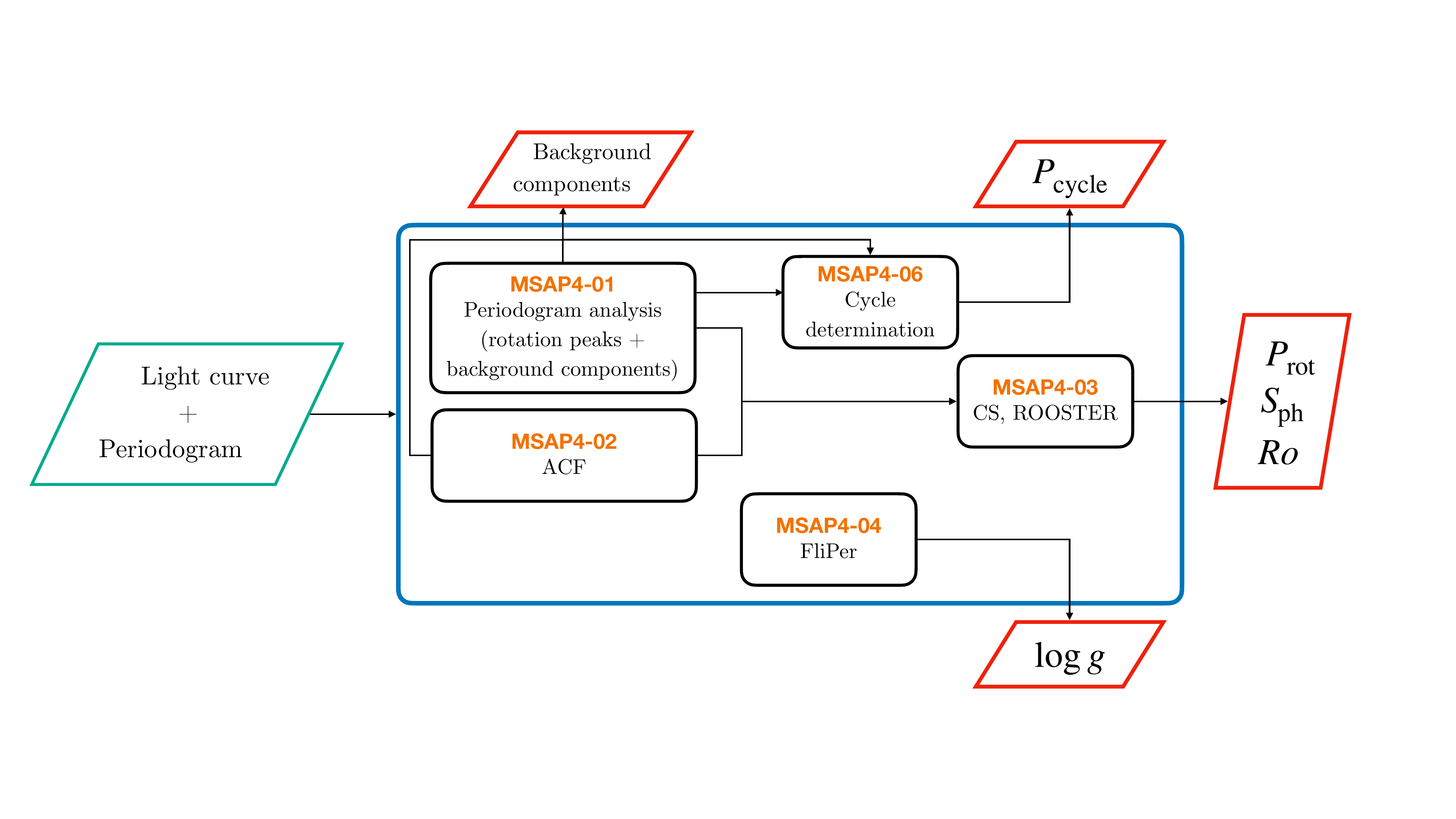}
    \caption{Flow diagram of the MSAP4 module.}
    \label{fig:msap4_diagram}
\end{figure}

The flow diagram of the MSAP4 module is shown in Fig.~\ref{fig:msap4_diagram}.
Using the stellar light curve and corresponding periodogram, MSAP4 will perform several analyses in order to provide values for the spectral background components \citep{Corsaro2014,Corsaro2015}, the logarithm of surface gravity, log g \citep{Bugnet2018}, the average surface rotation period, $P_\mathrm{rot}$, the photometric activity index, $S_\mathrm{ph}$ \citep{Mathur2014b}, the Rossby number, $Ro$ \citep{Noraz2022}, and long-term variability periods, $P_\mathrm{cycle}$. The demonstration code for MSAP4 is fully open-source\footnote{The source code can be downloaded here: \url{https://gitlab.com/sybreton/plato_rotation_pipeline}.} and modular, with dedicated documentation and tutorials\footnote{The documentation is hosted here: \url{https://plato-rotation-pipeline.readthedocs.io/en/latest/}.}. 

\subsection{Extracting rotation}

For stars ranging from M- to F- type, MSAP4 will measure surface rotation periods. 
To ensure robustness of the analysis, rotation-related parameters are
extracted with several methods. In MSAP4-01, Fourier analysis is performed with the generalised Lomb-Scargle \citep[GLS,][]{Zechmeister2009}. In MSAP4-02, a time series analysis is performed by computing the autocorrelation function \citep[ACF,][]{McQuillan2013b}
In MSAP4-03, the two methods are combined to enhance common periodicities, yielding the composite spectrum \citep[CS,][]{Ceillier2016}. These results are combined using the machine-learning methodology Random fOrests Over STEllar Rotation \citep[ROOSTER,][]{Breton2021}. 
In particular, ROOSTER will provide a reliability score for the measured period. 
$S_\mathrm{ph}$ and $Ro$ are finally computed using the measured $P_\mathrm{rot}$. 

\subsection{Looking for cyclic modulations}

In addition to surface rotation measurements, long-term variability in the light curves will be searched: solar-type magnetic activity cycle, but also shorter cycle such as Rieger-like
cycles \citep{Rieger1984}. To this purpose, measurements from the GLS, the ACF, and the $S_\mathrm{ph}$ modulation will be combined.

\section{Conclusions}

PLATO will enable to obtain new insights on the interaction between rotation and activity in main-sequence low-mass stars. Combined with the asteroseismic measurements that the mission will also perform, it will provide an unprecedented sample of stars with exquisite characterisation.
 
\begin{acknowledgements}
S.N.B, A.F.L., Se.M., E.C., and I.P. acknowledge support from PLATO ASI-INAF agreement n.~2015-019-R.1-2018.
R.A.G acknowledges support from PLATO and GOLF CNES grants. 
Sa.M.~acknowledges support from the Spanish Ministry of Science and Innovation (MICINN) with the Ram\'on y Cajal fellowship no.~RYC-2015-17697, grant no.~PID2019-107187GB-I00 and PID2019-107061GB-C66, and through AEI under the Severo Ochoa Centres of Excellence Programme 2020--2023 (CEX2019-000920-S).
A.R.G.S. acknowledges the support by FCT through national funds and by FEDER through COMPETE2020 by these grants: UIDB/04434/2020 \& UIDP/04434/2020. A.R.G.S. is supported by FCT through the work contract No. 2020.02480.CEECIND/ CP1631/CT0001.
\end{acknowledgements}

\bibliographystyle{aa}  
\bibliography{breton} 

\end{document}